\documentclass[a4paper]{spie}  
\usepackage[]{graphicx}
\usepackage{url}
\usepackage[latin1]{inputenc}
\usepackage[OT1]{fontenc}
\usepackage{multirow}

\title{The 2010 Interferometric Imaging Beauty Contest} 

\author{%
  Fabien Malbet\supit{a\dag}, %
  William Cotton\supit{b}, %
  Gilles Duvert\supit{a\dag}, %
  Peter Lawson\supit{c}, %
  Andrea Chiavassa\supit{d}, %
  \linebreak %
  John Young\supit{e}, %
  Fabien Baron\supit{f}, %
  David Buscher\supit{e}, %
  Sridharan Rengaswamy\supit{g}, %
  Brian Kloppenborg\supit{h}, %
  Martin Vannier\supit{i}, %
  Laurent Mugnier\supit{j}
\skiplinehalf
\supit{a}Lab.\ d'Astrophysique de Grenoble (LAOG), UMR 5571 Univ.\ J.\ Fourier/CNRS, BP 53, F-38051 Grenoble cedex 9, France;\\
\supit{b}National Radio Astronomy Obs., 520 Edgemont Road, Charlottesville, VA 22903, USA;\\
\supit{c}Jet Propulsion Lab., California Institute of Technology, Pasadena, CA 91109, USA;\\
\supit{d}Max-Planck-Institut für Astrophysik, Karl-Schwarzschild-Str. 1
D-85741 Garching, Germany; \\
\supit{e}Astrophysics Group, Cavendish Lab., JJ Thomson Avenue, Cambridge CB3 0HE, UK;\\
\supit{f}Univ.\ of Michigan, 941 Dennison Building, 500 Church Street, Ann Arbor, MI 48109, USA;\\
\supit{g}European Southern Obs.\, Casilla 19001, Santiago 19, Chile;\\
\supit{h}Univ.\ of Denver, 2112 East Wesley Ave, Room 211, Denver, CO 80208, USA;\\
\supit{i}Laboratoire H.\ Fizeau, Univ.\ de Nice-Sophia Antipolis, parc Valrose, F-06108 Nice, France;\\
\supit{j}ONERA/DOTA, BP 72, F-92322 Ch\^atillon cedex, France;\\
\supit{$\dag$}Jean-Marie Mariotti Center (JMMC), France;
}

\authorinfo{Send correspondence to F.\ Malbet\\
E-mail: Fabien.Malbet\@obs.ujf-grenoble.fr}

\begin{document} 
  \maketitle 

\begin{abstract}
  We present the results of the fourth Optical/IR Interferometry
  Imaging Beauty Contest. The contest consists of blind imaging of
  test data sets derived from model sources and distributed in the
  OI-FITS format. The test data consists of spectral data sets on an
  object "observed" in the infrared with spectral resolution. There
  were 4 different algorithms competing this time: BSMEM the Bispectrum
  Maximum Entropy Method by Young, Baron \& Buscher; RPR the Recursive
  Phase Reconstruction by Rengaswamy; SQUEEZE a Markov Chain Monte
  Carlo algorithm by Baron, Monnier \& Kloppenborg; and, WISARD the
  Weak-phase Interferometric Sample Alternating Reconstruction Device
  by Vannier \& Mugnier. The contest model image, the data delivered
  to the contestants and the rules are described as well as the
  results of the image reconstruction obtained by each method. These results
  are discussed as well as the strengths and limitations of
  each algorithm.

\end{abstract}



\section{INTRODUCTION}

This paper presents the results of the fourth Optical/IR
Interferometry Imaging Beauty Contest following similar contests in
2004\cite{2004SPIE.5491..886L}, 2006\cite{2006SPIE.6268E..59L} and
2008\cite{2008SPIE.7013E..48C}. These contests are intended to
encourage the development and distribution of techniques and software
for imaging optical/IR astronomical interferometric data. Due to the
phase noise introduced by atmospheric turbulence, phase information
measured on individual interferometric baselines is strongly
corrupted. However, the ``closure phase'' or ``bi-spectrum'' technique
uses the sums of measured phases around closed loops of baselines in
which all phase contributions other than those due to the target enter
twice but with opposite sign and therefore cancel. This technique has
been used to image data from COAST, NPOI, IOTA, ISI, CHARA and the
VLTI/AMBER interferometers.

The contest consists of imaging data sets of targets whose details are
unknown to the contest participants but are derived from models whose
Fourier transforms are sampled in the pattern of a typical
interferometer. The contest data is then the visibility square and
closure phases derived from this data set as would be measured by such
an array. In each contest, the targets have become more challenging.
In the 2010 contest, the team who has generated the data set has
mimicked AMBER/VLTI data allowing spectral imaging with the delivery
of spectrally-dispersed Fourier data cubes. The object is a supergiant
with a faint companion. This contest is being conducted by the Working
Group on Image Reconstruction of IAU Commission 54.

\section{Contest model, data and guidelines}

The 2010 challenge consisted in reconstructing the surface features of a
supergiant hosting a much fainter and unresolved companion located at
8 stellar radii.

\begin{figure}[t]
  \centering
 \parbox{0.4\hsize}{\centering
    \includegraphics[width=0.3\textwidth]{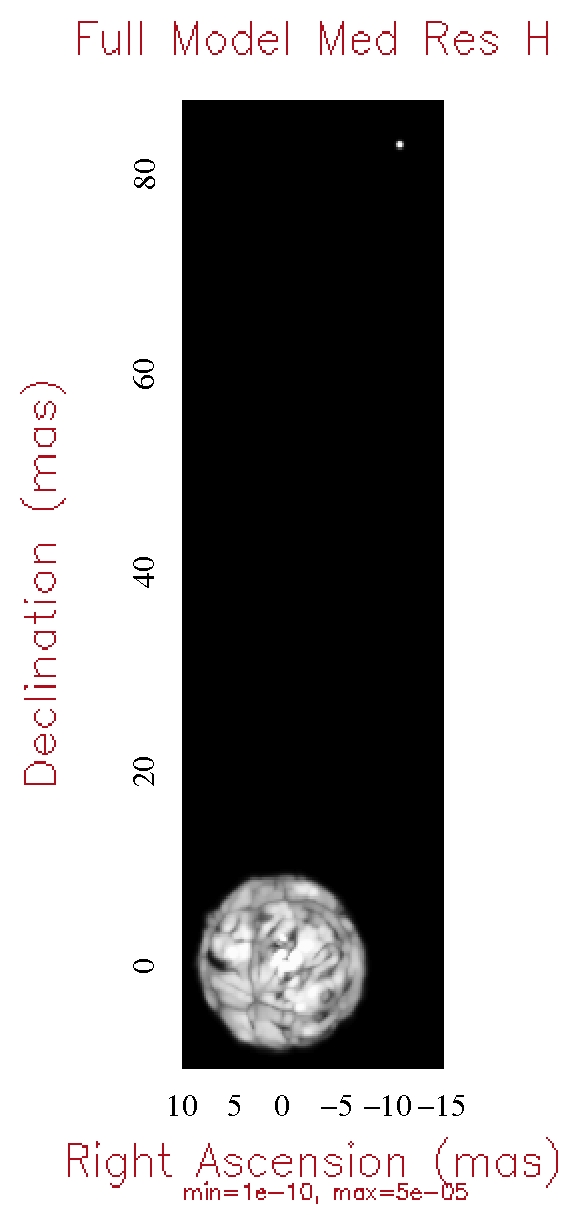}}
  \parbox{0.4\hsize}{\centering
    \includegraphics[width=0.3\textwidth]{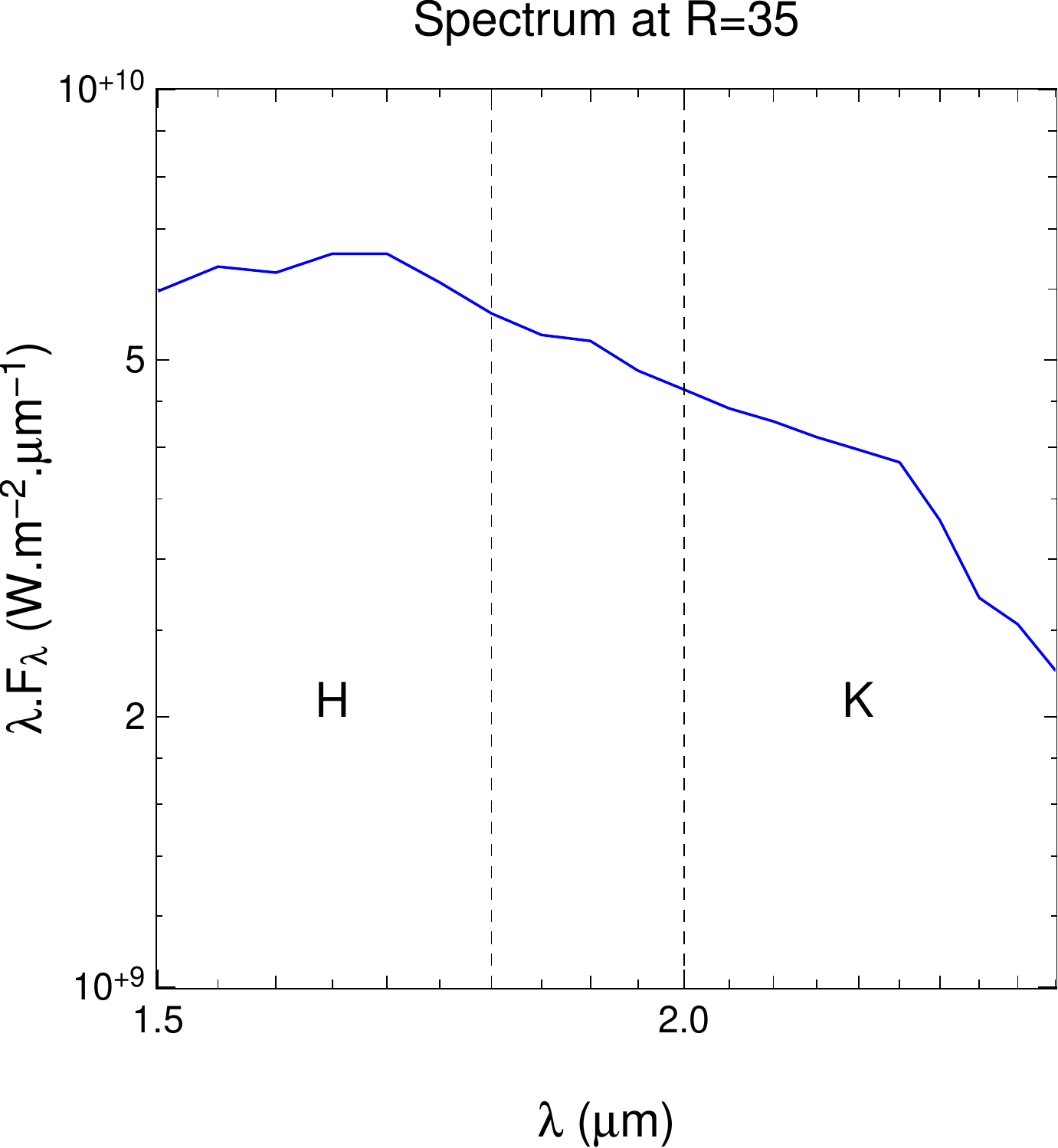}\\
    \includegraphics[width=0.3\textwidth]{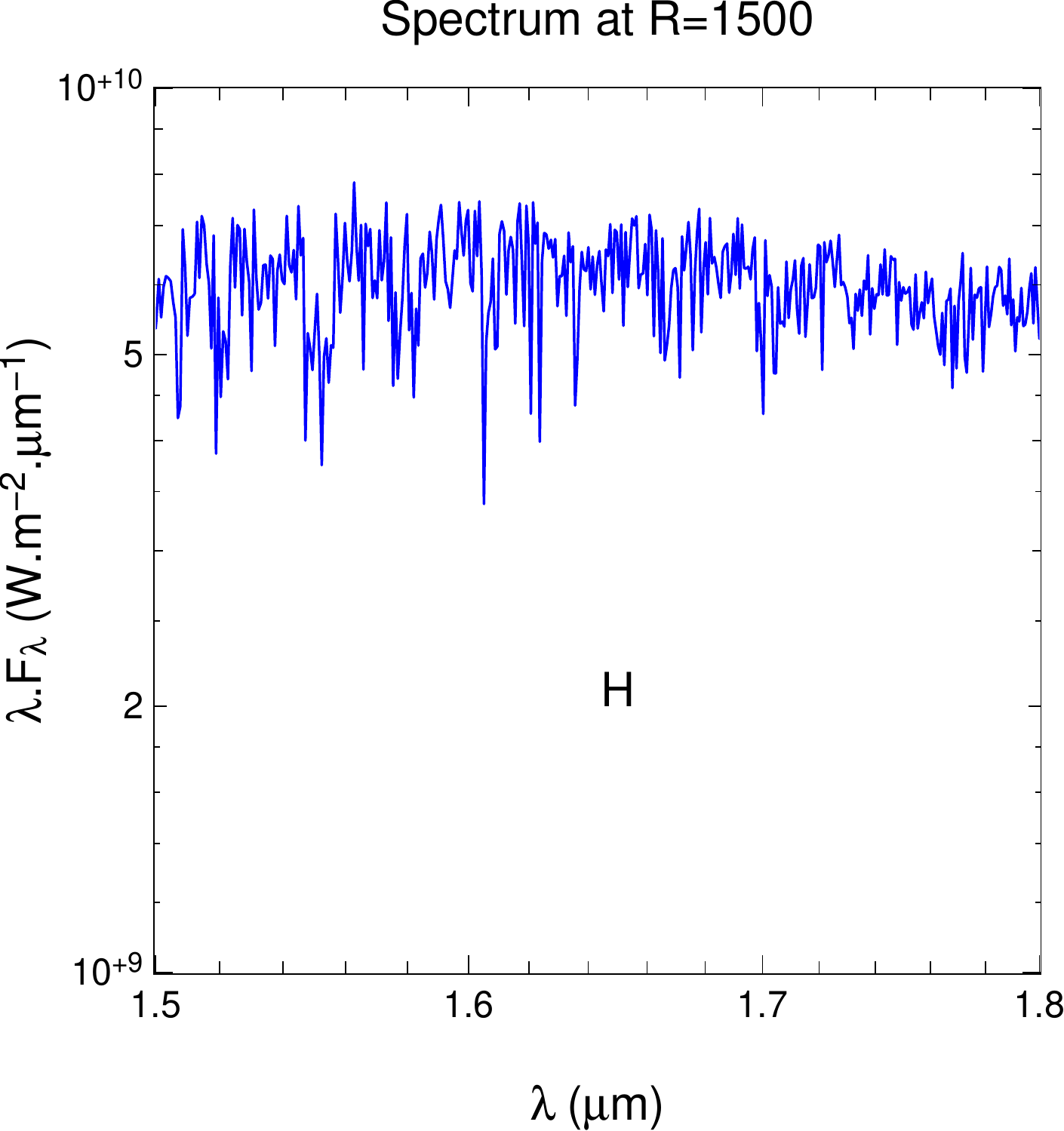}
  } \medskip
  \caption{Left: Original image of the first plane of the data cube
    used to produce the 2010 Beauty Contest. It consists in a stellar
    surface of a supergiant (as in Chiavassa et al.\
    2009\cite{2009A&A...506.1351C}), set at the angular dimension and
    location of the supergiant $\gamma$ Cru, and a faint companion
    located 80\,mas to the North-West averaged over the MRH data
    cube. Right: the spectrum of the supergiant in low resolution
    (top) and in medium spectral resolution (bottom).}
  \label{fig:orig-image}
\end{figure}

\begin{figure}[p]
  \centering
  \includegraphics[width=0.4\hsize]{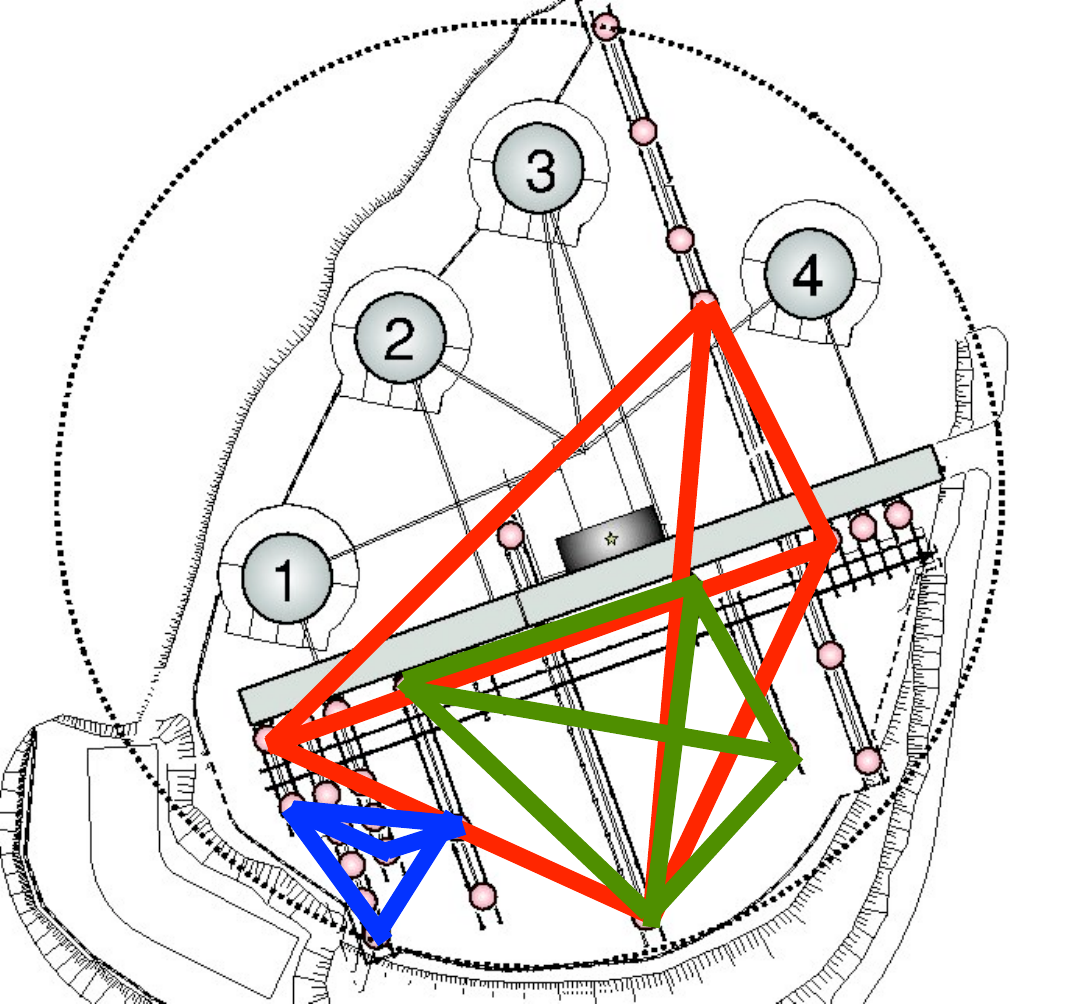}
  \includegraphics[width=0.55\hsize]{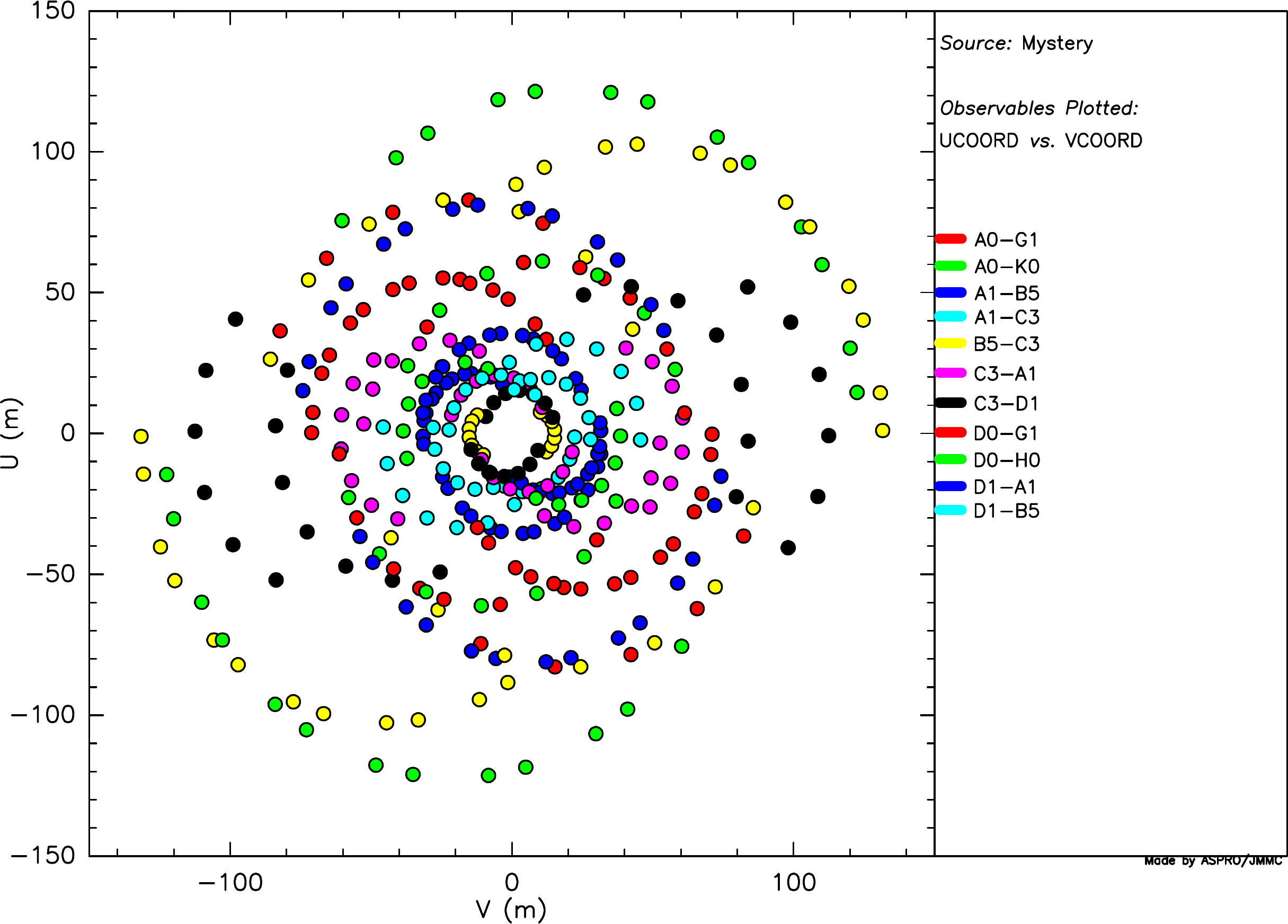}
 \caption{Baseline configuration at VLTI (left) used for the contest
    and the resulting $(u,v)$ plane coverage (right).}
  \label{fig:uvcov}
\end{figure}

\begin{figure}[p]
  \centering
  \begin{tabular}{cc}
    \parbox{0.45\hsize}{\centering \includegraphics[width=0.7\hsize]{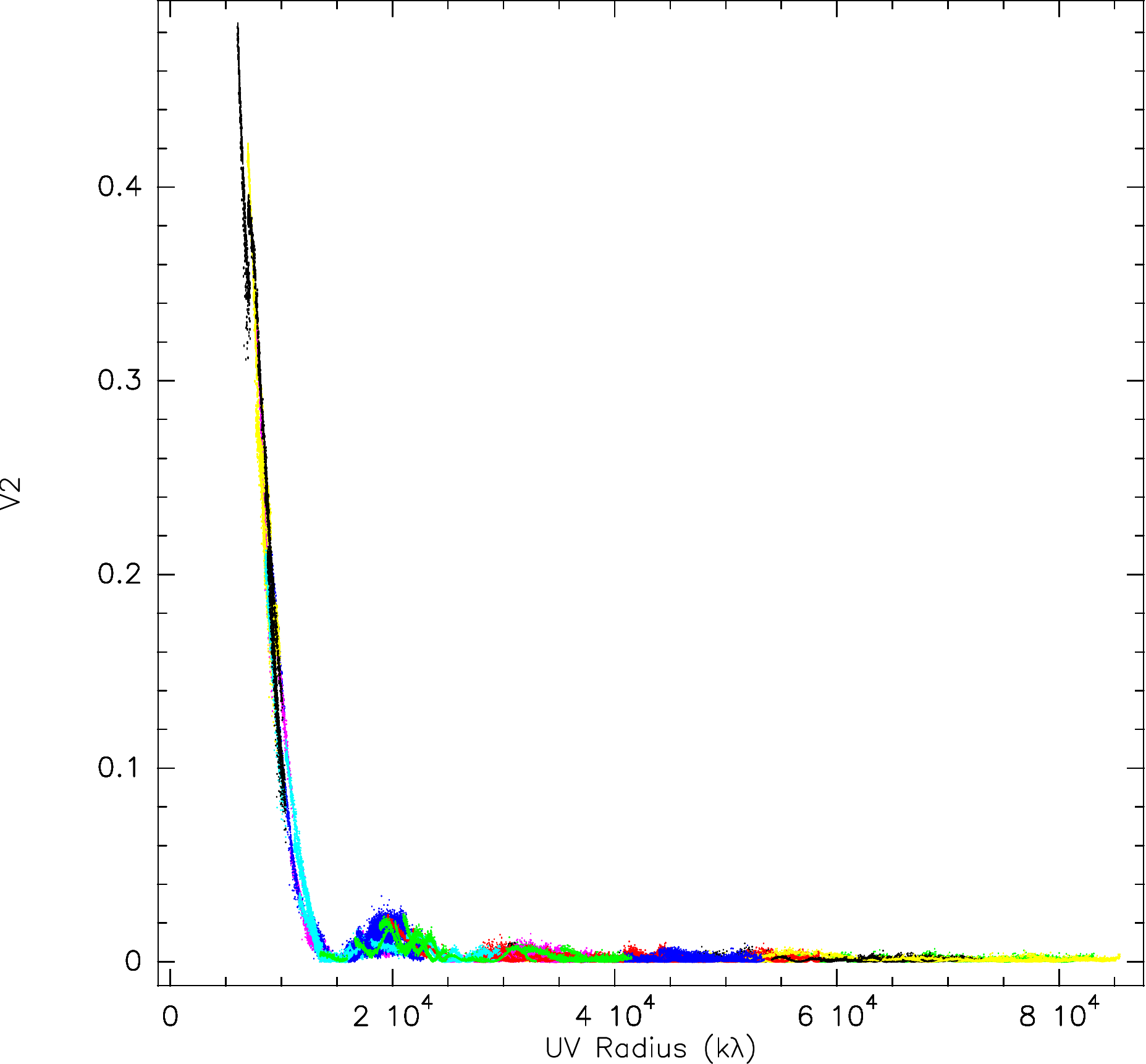}} &
    \parbox{0.45\hsize}{\includegraphics[width=\hsize]{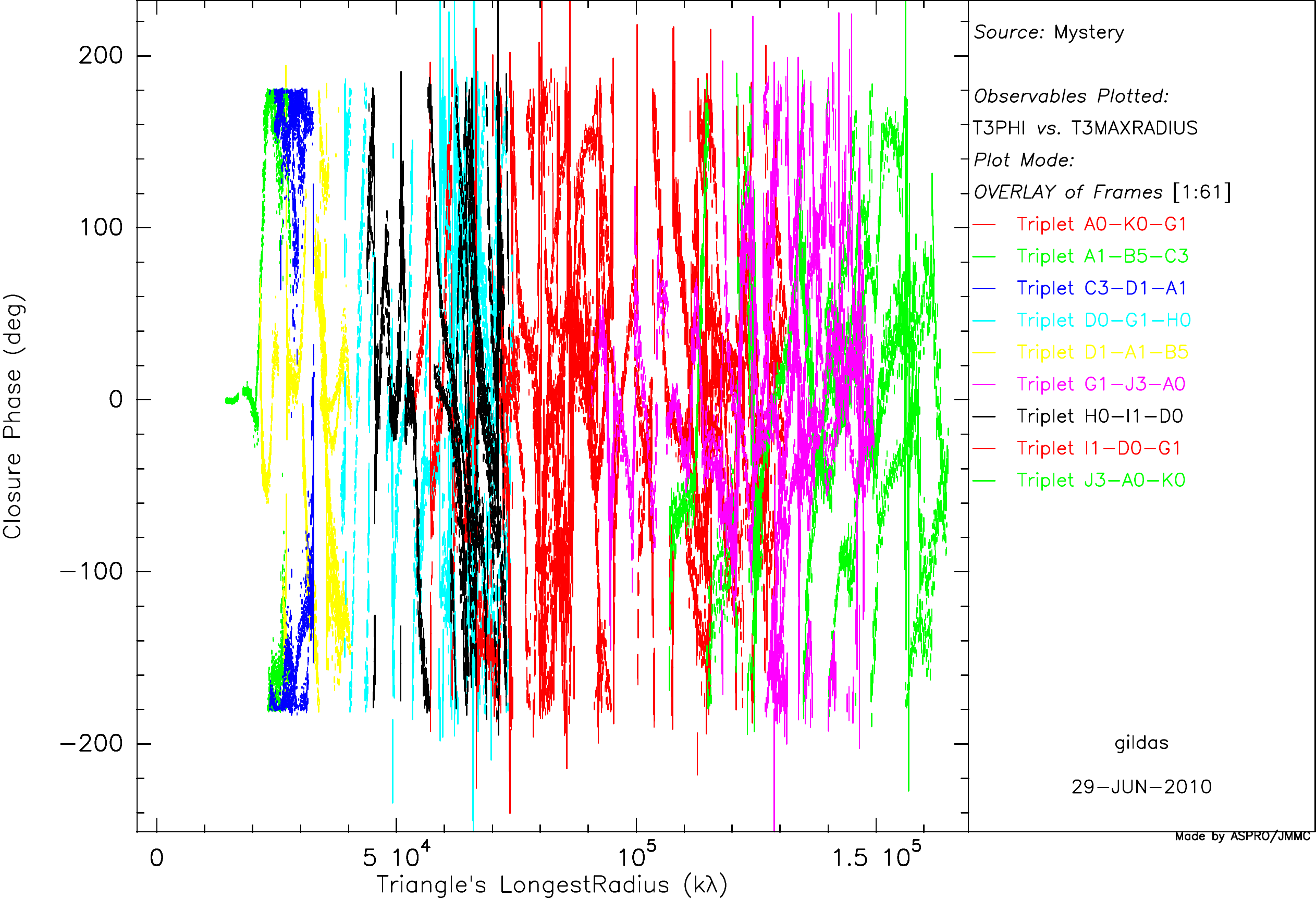}} \\
    \parbox{0.45\hsize}{\includegraphics[width=\hsize]{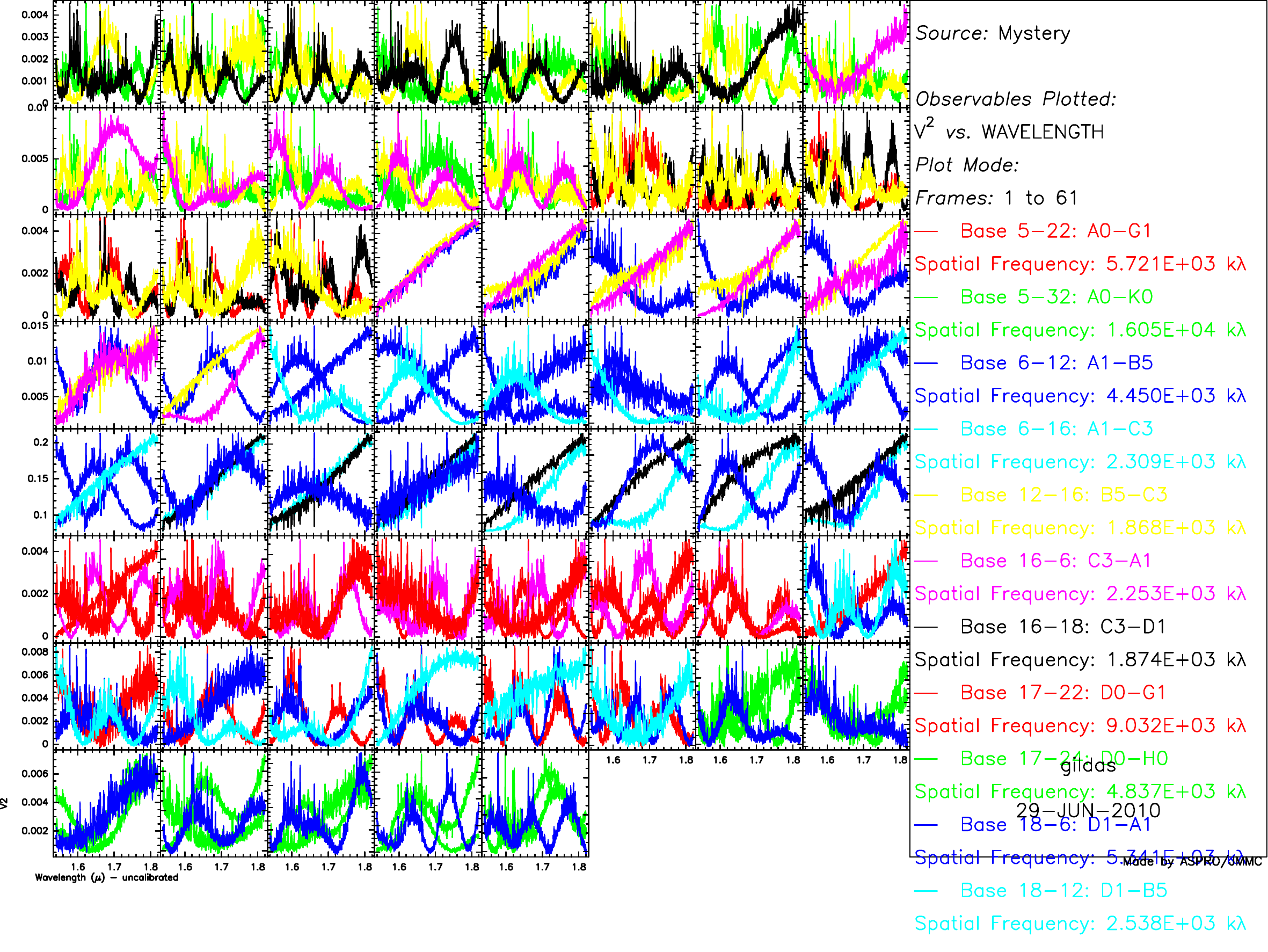}} &
    \parbox{0.45\hsize}{\includegraphics[width=\hsize]{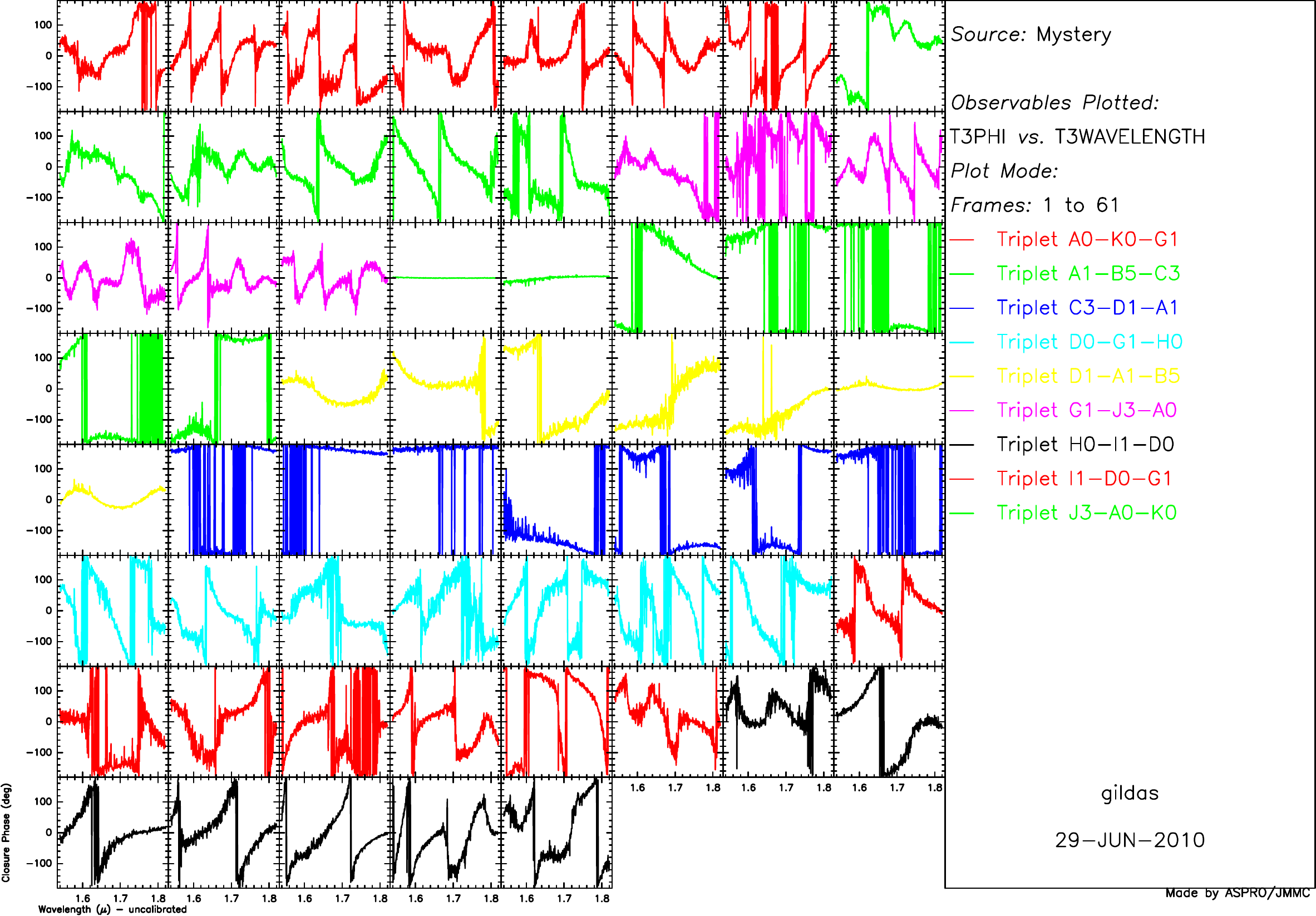}} \\
 \end{tabular}
  \caption{Interferometric data delivered to the contestants. Left:
    squared visibilities; right: closure phases. Top: wrt.\
    the spatial frequencies; bottom: wrt.\ the wavelength.}
  \label{fig:oi-data}
\end{figure}

\subsection{The original data}

The original image comes from the work of Chiavassa et
al. (2009)\cite{2009A&A...506.1351C} who computed intensity maps at
different wavelengths (corresponding to broadband filters $H$ and $K$)
with the radiative transfer code OPTIM3D from snapshots of the 3D
hydrodynamical simulation (code CO$^5$BOLD) of a red supergiant star.
They used in this work a model with stellar parameters close to those
of Betelgeuse \cite{2005ApJ...628..973L}: a 12 $M_{\odot}$ stellar
mass, a luminosity averaged over spherical shells and time of
$L$=93000$\pm$1300~$L_{\odot}$, an effective temperature of
$T_{\rm{eff}}$=3490$\pm$13~K, a radius of $R$=832$\pm$0.7~$R_{\odot}$,
and surface gravity log($g$)=-0.337$\pm$0.001. The numerical
resolution is 235$^3$ with a grid spacing of 8.6~$R_{\odot}$.  The
model of this M0 40\,mas supergiant has been scaled to a
$\approx$20\,mas star and set at the location of $\gamma$ Cru whose
angular size matches and could arguably have such features, and is
conveniently imageable by the VLTI.

To add a bit of spice, we also put a faint companion at distance
$\rho=84.28$\,mas at $PA=261.31$\,degrees. This separation is only 8
stellar radii of the stellar source, hence resolving the companion is
well beyond the capabilities of adaptive optics instruments like NACO.
The companion is 5 magnitudes fainter than the star. This translates
to an A0V-like star or so in terms of stellar type. It should be
visible in the image reconstruction at least the medium resolution $H$
images, because of the small ripple it gives at very low V2.  Also, it
would be interesting to see if phase closure 'nulling' signatures like
those described in Chelli et al.\ 2009\cite{2009A&A...498..321C}
permit to retrieve the companion spectrum.

Figure \ref{fig:orig-image} displays the original image with the
stellar surface in the South East corner and the unresolved companion
in the North-West region.

\subsection{Interferometric data}

\begin{figure}[t]
  \centering
  \includegraphics[width=0.7\hsize]{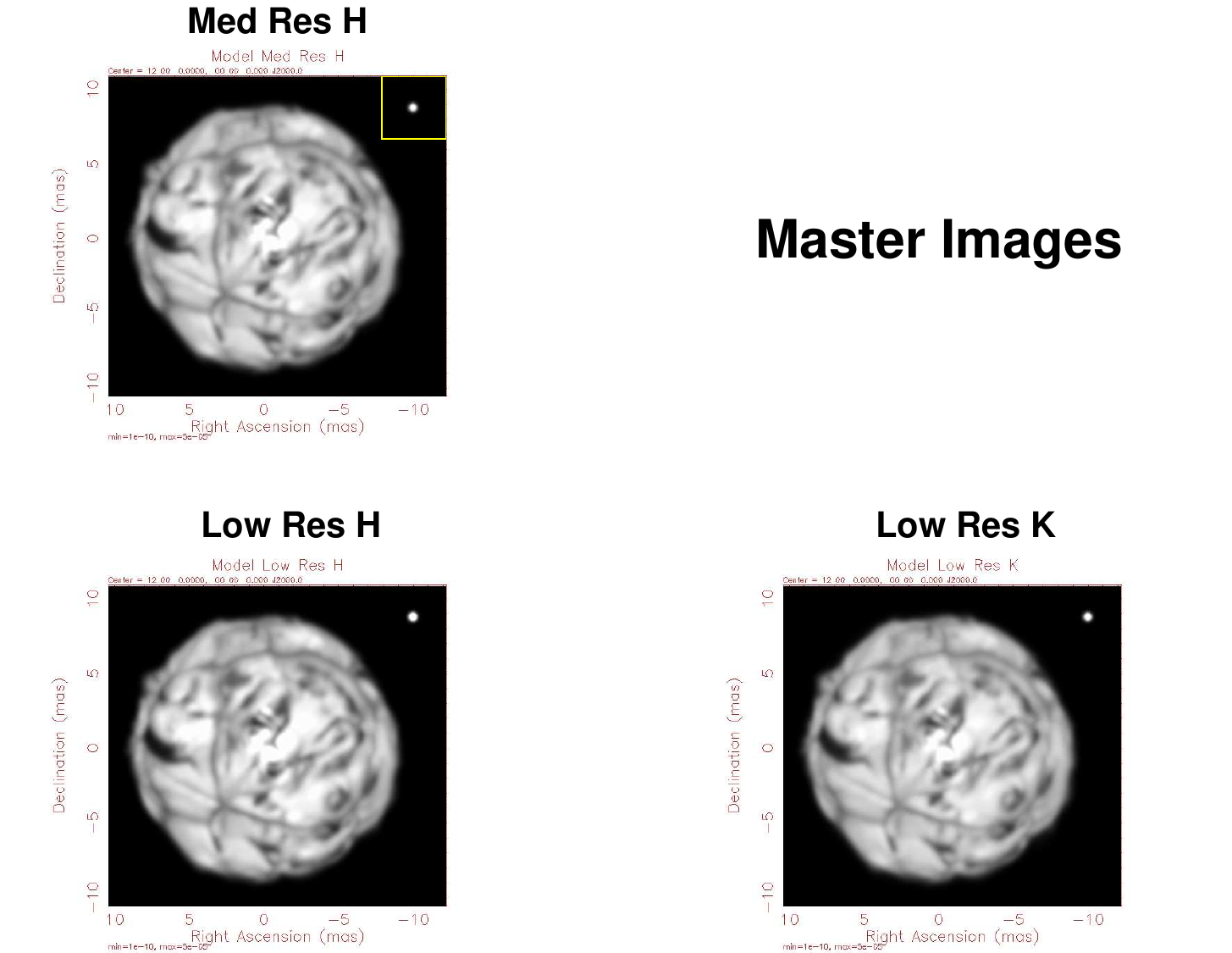}
  \caption{Gray master images used for the comparison. The upper left
    is the gray image resulting from the medium resolution data cube
    in $H$ band, the left lower image from the low resolution in $H$
    band and the right lower image to the low resolution in $K$
    band. In the upper right of each image the inset represents the
    part of the image where the unresolved companion is located. It
    has been moved closer to the stellar surface in order to have
    better details of both components.}
\label{fig:master}
\end{figure}

The simulations used here have been made with the \texttt{ASPRO}
package\cite{2002SPIE.4844..295D}. \texttt{ASPRO} simulates the
observation by an interferometer of a science astronomical object, at
one or several times and delivers simulated interferometric
observables in the OIFITS format\cite{2005PASP..117.1255P}. The
interferometric chain is modeled as the combination of an
interferometer infrastructure and focal instruments.  The
interferometer infrastructure comprises the telescopes, delay lines,
tip-tilt correctors, adaptive optics, and fringe trackers. It adds
geometrical requirements such as the positions and sizes of the
telescopes apertures at the time of observation with respect to the
position of the science object and the geometrical delays thus induced
between each pair of telescopes. In addition, it includes
environmental constraints such as the atmospheric seeing, the
different horizons seen by telescopes and technical limitations
(limits on the delay-line strokes, flux dependence of active optical
elements...).

Two observations of the synthesized $\gamma$ Cru object were simulated:
a $H$-band medium spectral resolution (MRH) at $R=1500$, and a
simultaneous H- and K-band low spectral resolution (LRHK) at $R=35$
with the 1.8-m telescopes of VLTI, which are meant for imaging. The $(u,v)$
coverage for imaging has been simulated for the 3T instrument AMBER,
using series of triplets taken in the quadruplets proposed by ESO/VLTI
for their period 87: A0-G1-K0-J3, A1-B5-C3-D1, D0-G1-H0-I1. This
configuration can be achieved in 3 days, one day per quadruplet by
switching AMBER beams every 30 minutes between 3 triplets
of the quadruplets: A0-G1-K0, A0-J3-K0, A0-G1-J3 A1-B5-C3, A1-C3-D1,
A1-B5-D1 D0-G1-H0, D0-H0-I1, DO-G1-I1.

The basic idea was to test the reconstruction methods for
spectrally-dispersed data. To enable a channel-per-channel
reconstruction, the signal-to-noise ratio per spectral channel needed
to be good, thus the object bright. The size of the star and the
principal features could be retrieved with a 'grey' image
reconstruction, or by binning a number of spectral channels. It could
be possible also to use the 'grey' or 'binned' images as a starter for
a finer, channel by channel, reconstruction or by forcing the
reconstruction support inside a 20 mas stellar disc.

\subsection{Contest rules}

This year, there was two sets of judgings: 
\begin{enumerate}
\item one for "gray" images (it was agreed that "gray" meant uniform
  weighting of spectral channels) derived from the provided datasets
  assuming a constant target structure with wavelength:
  \begin{enumerate}
  \item one derived from all the channels in the
    \texttt{Mystery-Med\_H.oifits.gz} data set,
  \item one from the first 10 (short wavelengths) channels of the
    \texttt{Mystery-Low\_HK.oifits.gz} data set and
  \item the third from the second 10 channels of the
    \texttt{Mystery-Low\_HK.oifits.gz} data;
  \end{enumerate}
  and %
\item a second for a "spectral" image allowing channel to channel
  differences.
\end{enumerate}
Submissions could be sent for either or both categories. Spectral
image submissions could be either as a single three dimensional FITS
cube with 512 planes or as 512 separate single plane FITS images.

\section{Contest submission}

\subsection{BSMEM}
\subsubsection*{by Young, Baron and Buscher (University of Cambridge)}

\begin{figure}[t]
  \centering
  \includegraphics[width=0.7\hsize]{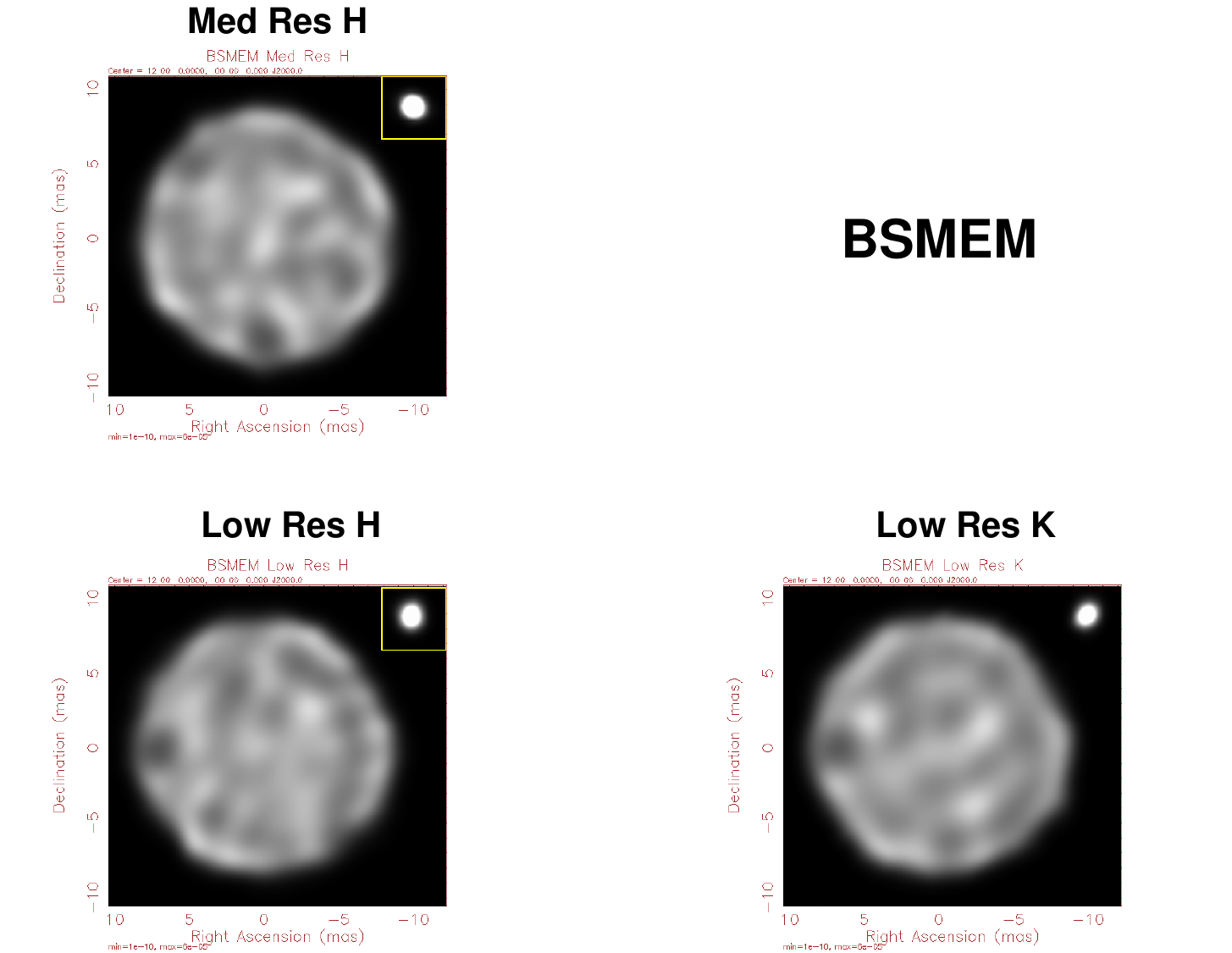}
  \caption{Gray scale images of BSMEM images as used in the comparison. Display is the same as in Figure \ref{fig:master}.} 
\end{figure}

The BSMEM (BiSpectrum Maximum Entropy Method) software was first
written in 1992 to demonstrate image reconstruction from optical
aperture synthesis data, and has been extensively enhanced and tested
since then. It applies a fully Bayesian approach to the inverse
problem of finding the most probable image given the evidence, making
use of the Maximum Entropy approach to maximize the posterior
probability of an image. BSMEM is available free-of-charge to the
scientific community on submission of the academic licence agreement
at \url{http://www.mrao.cam.ac.uk/research/OAS/bsmem.html}.

BSMEM uses a trust region method with non-linear conjugate gradient
steps to minimise the sum of the log(likelihood) (chi-squared) of the
data given the image and a regularization term expressed as the
Gull-Skilling entropy $\sum_k (I_k - M_k - I_k \log(I_k/M_k))$. The
model image $M_k$ is usually chosen to be a Gaussian, a uniform disk, or
a delta-function centered in the field of view, which conveniently
fixes the location of the reconstructed object (the bispectra and
power spectra being invariant to translation).  This type of starting
model also acts as a support constraint by penalising the presence of
flux far from the centre of the image. An important advantage of BSMEM
is the automatic Bayesian estimation of the hyperparameter alpha that
controls the weighting of the entropic prior relative to the
likelihood.  BSMEM can also perform a Bayesian estimation of missing
triple amplitudes and their associated errors from the powerspectra
data - it was necessary to use this capability for the contest data.

We found that BSMEM was very slow (runtimes upwards of 24 hours on a
standard PC) to converge to a well-fitting (reduced chi-squared $< 5$)
solution when given data from multiple spectral channels and an
uninformative prior - probably in large part due to the
wavelength-dependence of the object. Thus we followed the approach of
finding an image whose low spatial frequencies were compatible with
the data, and reconstructing each spectral channel separately using
this as the model image. The reconstructed images from each spectral
channel were then averaged to generate the three gray images specified
by the contest organizers.

In particular, we found that BSMEM did eventually converge using
low-spatial-frequency data (baselines up to 60m) from the first 8
spectral channels of the low-spectral-resolution dataset. The
resulting image was convolved with a Gaussian with FWHM equal to the
finest fringe spacing and thresholded to remove features judged to be
noise. This convolved image was used as the model image for the
subsequent reconstructions of individual spectral channels, which
typically converged after 70 to 250 iterations (of order 30 minutes on
a standard PC). To generate the gray submissions, the spectral
channels were averaged with uniform weighting, and pixels below a
specified threshold in the average image were set to zero. The
threshold value was chosen to remove most of the obvious circumstellar
artefacts, each of which typically appeared in the one spectral channel
only.

In the submitted FITS images, North is up and East is to the left when
the standard display convention for FITS images is followed. I also
attach a reconstruction (0.5\,mas/pix) of the contest binary to confirm
this. BSMEM does not impose any resolution on the reconstructed image,
except for the pixel size which was chosen to be 0.4\,mas for the
mystery object (~6 pixels per fastest fringe). The size of the smallest
believable features in the reconstructions is ~0.8\,mas, so you may wish
to use this resolution for comparison with the truth images.

We are confident the following features are real:
\begin{itemize}
\item The "stellar" disk and its limb-darkening
\item The compact companion close to the north edge of the maps
\item The more prominent stellar surface structures (in particular,
  the central bright spot and the two prominent bright spots in the
  northern hemisphere, the dark area close to the southern disk edge,
  and probably the dark area close to the eastern disk edge)
\item The changes in the contrast of the stellar surface structures
  with wavelength
\end{itemize}

\subsection{RPR}
\subsubsection*{by Rengaswamy (European Southern Observatory)}
\begin{figure}[t]
  \centering
  \includegraphics[width=0.7\hsize]{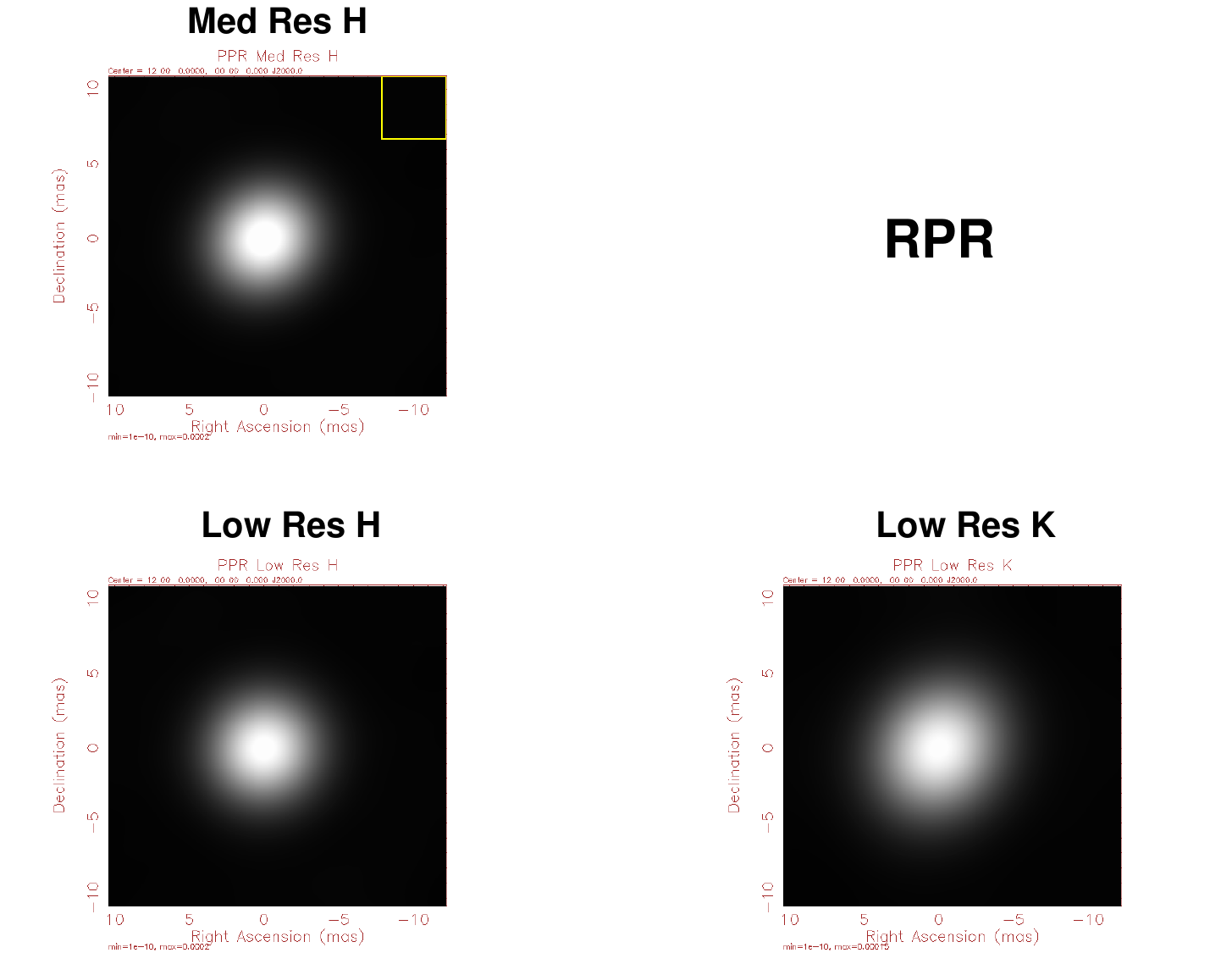}
  \caption{Gray scale images of RPR images as used in the comparison. Display is the same as in Figure \ref{fig:master}.} 
\end{figure}

For the contest data, images of size $512\times512$ pixels were
synthesized over a field-of-view of $0.2048\times0.2048$\,arcsec, in
each of the spectral channels (10 each in LRH and LRK and, 512 in
MRH). For contest \#1, the individual images were averaged over the
spectral line channels to obtain 3 2-D images. For contest \#2, the
individual images are stored in 10 data cubes.

I think, the mystery object is a binary; at least it is the most
obvious feature; there could be some extended background features but
they are not reconstructed clearly. There is a central bright object,
which is well resolved disk, with angular size of about 14\,mas. The
secondary is at a distance of $17.3\pm0.56$\,arcsec, with position angle of
153\,deg (east of north). The secondary is fainter by a factor of
40. There could be some extended background in both the images, but
they are not reconstructed.

\subsubsection*{Procedure (software)}

\begin{enumerate}
\item Visibility amplitude was obtained directly from $V^2$ data.
\item Bispectra were estimated for a ``model'' source. Then observed
  bispectra were substituted in place of the model bispectra and a
  phase map was obtained using a recursive reconstruction method.
\item Complex visibilities (obtained from step 1 and 2) were weighted
  according to their signal-to-noise ratio, averaged at grid points as
  and when required, weighted by a tapering function (Hanning
  function) and then Fourier inverted.
\item The map is then CLEANED over the central region and the CLEAN
  residuals are added.
\end{enumerate}

\subsubsection*{Known issues} 

Perhaps due the basic assumptions of ``CLEAN'', the background is not
reconstructed. The individual reconstruction in spectral channels show
some 180 degrees ambiguity for the secondary- this could be possibly because
of incorrect phase information.

\subsubsection*{Note}

The reconstructed images contained several bright
points, which, I suspected, are arising from amplitude errors. So I
did a ``PHASE ONLY RECONSTRUCTION'', and found that these bright
features (4 of them) go away. Thus, for the results I presented here,
I did phase only reconstructed images (to avoid artifacts). I did not
use visibility amplitude and differential phases. I used only closure
phases.

\subsection{SQUEEZE}
\subsubsection*{by Baron, Kloppenborg and Monnier (University of
  Michigan and University of Denver)}
\begin{figure}[t]
  \centering
  \includegraphics[width=0.7\hsize]{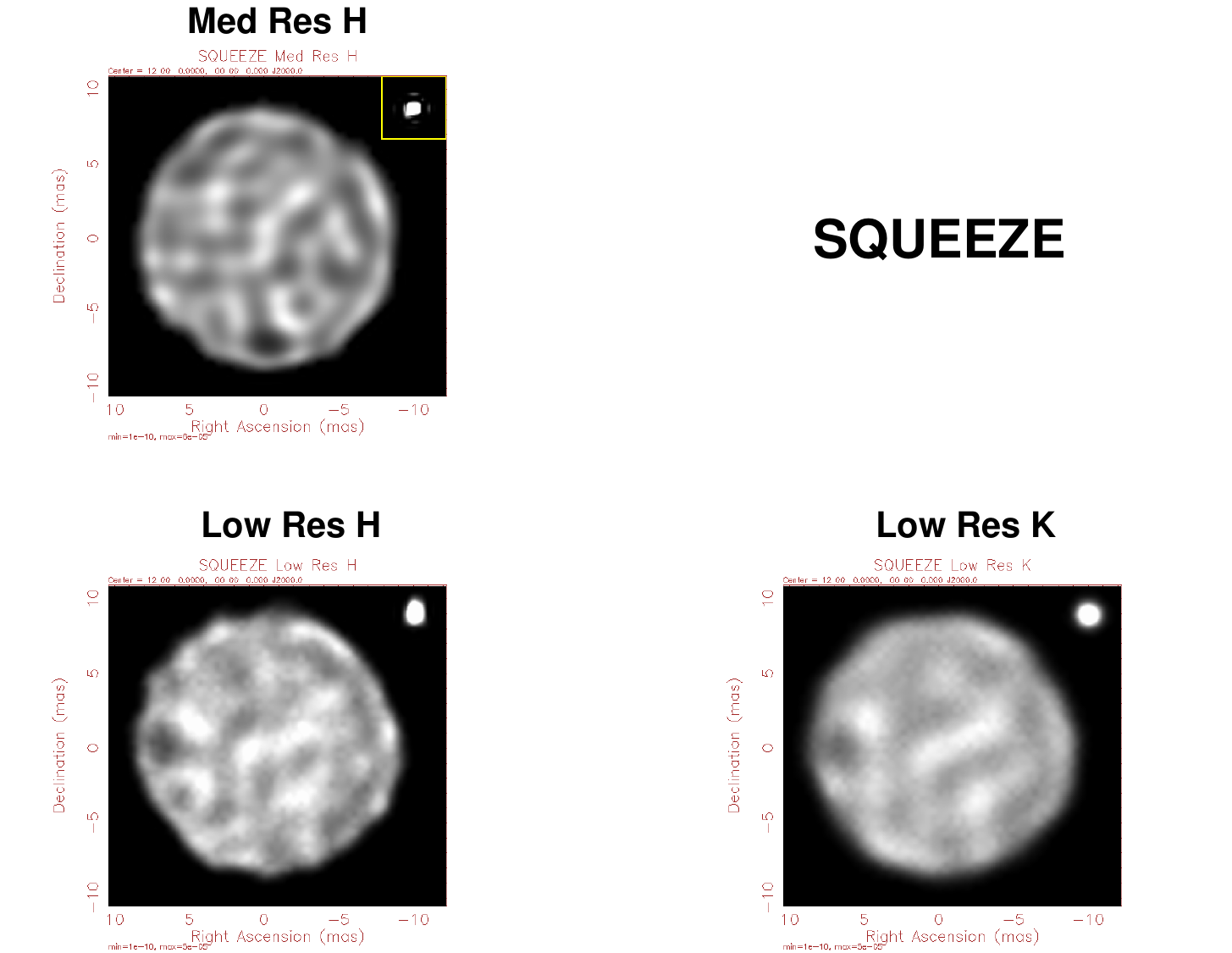}
  \caption{Gray scale images of SQUEEZE images as used in the comparison. Display is the same as in Figure \ref{fig:master}.} 
\end{figure}

SQUEEZE is a new software developped by Fabien Baron and John Monnier
at the University of Michigan, with the collaboration of Brian
Kloppenborg from the University of Denver. It is the spiritual
sucessor of MACIM, but also benefits from code exchange with the GPAIR
project and the model-fitting code Fitness. Both SQUEEZE and GPAIR are
in their infancy, and are described for the first time in these
proceedings. SQUEEZE uses a Markov Chain Monte Carlo engine, and
relies on simulated annealing and parallel tempering to sample the
posterior probability distribution of the image and find its global
maximum. Contrary to other software, it does not explicitly search
for the most probable "mode" image, but collects/averages statistics
around it. A secondary gradient-based engine is being added to
precisely determine the mode image. Finally a complementary module
currently allows simultaneous model-fitting and image reconstruction.
Novel capabilities for interferometry such as wavelet regularization
within the compressed sensing framework and imaging on spheroids are
also being implemented.

For this contest we uses the quantified Shannon entropy for our first
trials to obtain a draft image. We then convolved it by a PSF
corresponding to a 50-m baseline size to generate a prior image, which
we subsequently used within a Poisson entropy. Quantification allowed
the full process to be free of hyperparameters.

Our submission consists only of the grey images as we lacked the time
to implement an optimal algorithm for wavelength-dependent images
(full cube reconstruction with SED and differential visibility code)
compatible with our engine. For the low spectral resolution images,
convergence ("burn-in") to the lowest chi-squared was achieved for
$512\times512$ in less than a minute, after that the image was created
in about 30 minutes on a typical PC, by averaging 4000 posterior
samples. Reconstructing the medium spectral resolution image with this
number of samples was a much more computer intensive task, requiring
about two hours, but uneventful otherwise.

As an experiment, simultaneous imaging/model-fitting of a
bandwidth-smeared uniform disk allowed us to determine the position of
the object lying 80 degrees East of North as (-12.5, 84.7)\,mas, that
its flux represents about 2\% of the flux of the image, and that it
seems to be a point source in H (diameter $< 0.4$\,mas) and barely
resolved in K ( diameter $< 0.8$\,mas). The $64\times64$ images of the stellar
surface obtained in this "imaging/fitting" mode were obtained in less
than 2 minutes, were slightly better than the "full imaging" we
finally submitted. However we felt this would depart from the pure
imaging spirit of the contest, imaging a point source being for us
part of the challenge. Note that forcing a different regularization on
the point source location would have been possible but was not
attempted.

What we think is real is the limb-darkened stellar surface of the
primary, including the main discernable features (dark spot on the
East side for the low resolution datafiles and South for the medium one,
brighter features in the center, darker South West region). The
differences between the $H$ and $K$ images (apart from the visible effects
of the lower resolution in $K$) are subtle but are probably real
too. Finally the companion (Jupiter ?) has no discernable features but
is obviously real. The point source object is not as optimally
represented as we would wish without the model-fitting mode. The lower
dynamical range allowed by the engine (flux quantification and
propagation of flux quanta) does not easily allow us to represent such
a strong point source. Finally, exploring the posterior distribution
reveals unlikely features such as slight ejection in $H$ west of the
star or an ejection disk 50\,mas around the star which we do not
believe are part of the truth image and attribute to noise fitting
regions.

\subsection{WISARD}
\subsubsection*{by Vannier and Mugnier (Universit\'e of Nice and ONERA)}
\begin{figure}[t]
  \centering
  \includegraphics[width=0.7\hsize]{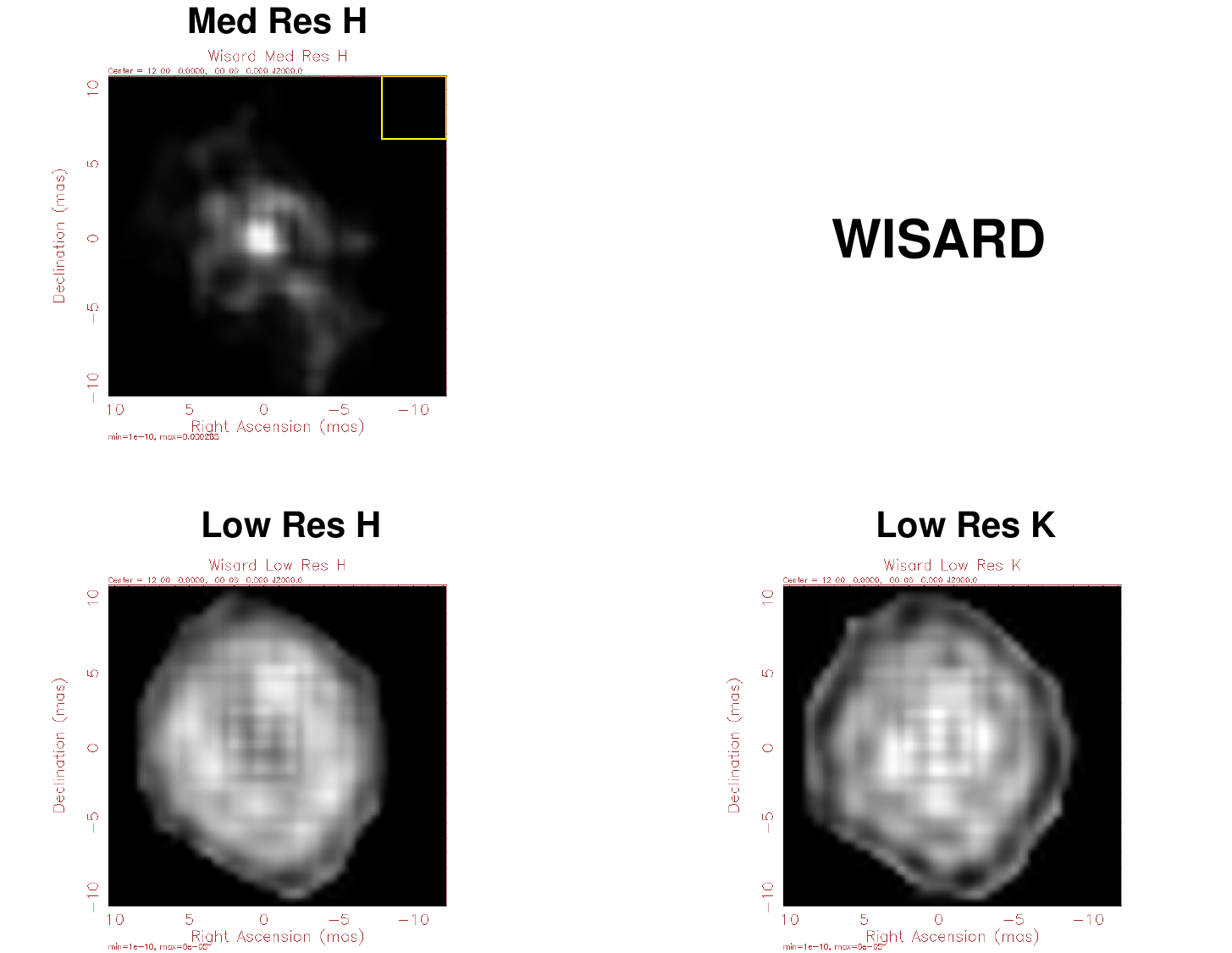}

  \caption{Gray scale images of WISARD images as used in the comparison. Display is the same as in Figure \ref{fig:master}.} 
\end{figure}

WISARD is an image reconstruction method developed
by Meimon et al.\cite{2004SPIE.5491..909M,2008JOSAA..26..108M} and designed to process optical long
baseline interferometry data, namely squared visibilities and closure
phases. It draws from self-calibration algorithms in
radio-interferometry\cite{1981MNRAS.196.1067C}, but uses a more
precise approximation of the data model\cite{2005JOSAA..22.2348M}
than first attempts\cite{1998JOSAA..15..811L}.  WISARD relies
on an explicit modeling of the missing phase information and allows
one to obtain a convex intermediate image reconstruction criterion.
The adopted approach consists in jointly finding the object and a
phase vector, corresponding to phase components in the null space of
the closure operator, that best fit the data.

WISARD also draws from Bayesian inversion methods in the sense that it
incorporates prior information on the sought object through a
regularization metric. Three regularization metrics are currently
implemented: (a) a basic Power Spectral Density-based regularization,
often used in single-aperture imaging\cite{1998ApOpt..37.4614C},
which was used in the 2004 IBC\cite{2004SPIE.5491..886L,
  2004SPIE.5491..886L}; (b) a non-quadratic, spike-preserving
regularization\cite{2008JOSAA..26..108M}, appropriate for objects
that mix point-like features and smooth areas; (c) a so-called
soft-support regularization, proposed specifically for long baseline
interferometry\cite{2008ISTSP...2..767L} in order to enforce
compactness of the sought object and thus interpolate missing
frequencies. WISARD is a free (open source) software developed in IDL
and has been used to process several astronomical
datasets\cite{2008A&A...485..561L, 2009A&A...508..923H,
  2009ApJ...707..632L}.

Concerning the images of the 2010 Beauty Contest, the FOV for the reconstructed $K$-band
image is 89 mas with $161\times161$ pixels (ie. a pixel size of
$0.553\times0.553$\,mas$^2$/pix) and whereas the FOV and number of pixels for
the $H$-band reconstructed images (both grey-LR and chromatic-MR) are
respectively 107.4\,mas and 193\,mas, which means almost the same
pixel size ($0.557\times0.557$\,mas$^2$/pix). No rotation was applied
to the images, they are oriented the standard way, North-East.

Concerning which features are interpreted as 'real', We certainly think
that the central part is real, with a peak surrounded by a fairly
compact (although not very regular) structure, oriented mainly from
top-left to bottom-right of the image.  We tend to think as well that
most of the irregular structures surrounding this central compact part
belong to a "ring" or a "layer", of diameter approximately half of the
larger field, i.e. about 50 mas.  As for the outest, and most
irregular structures, they might well be reconstruction artifacts.
We have had troubles regularizing such artifacts properly, given the very
large computat
ion time (lots of visibilities, large field,... and many
spectral channels).

\section{Comparison metrics and contest result}

\begin{figure}[t]
  \centering
  \includegraphics[width=0.7\hsize]{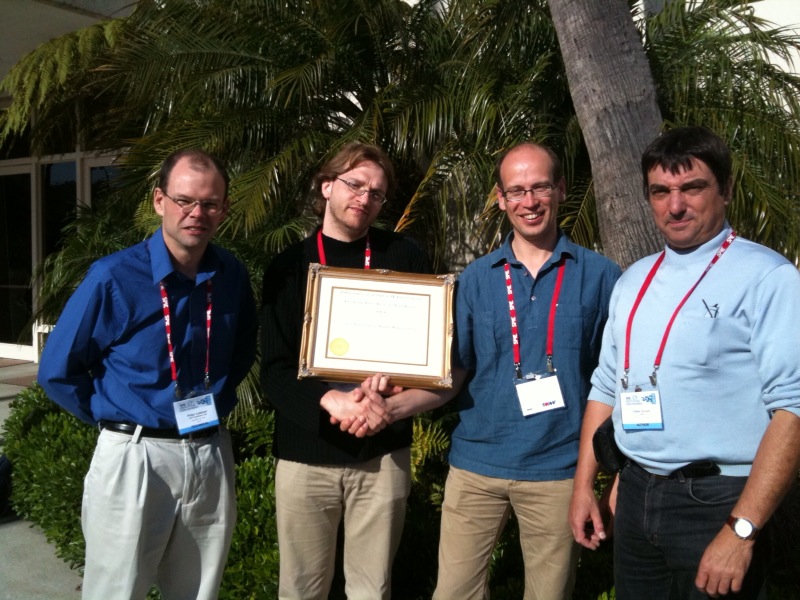}
  \caption{The 2010 Interferometry Imaging Beauty Contest jury
    presenting the award for the winning entry to the representative
    of the BSMEM team. From left to right: P.\ Lawson, F.\ Baron, F.\
    Malbet, G.\ Duvert.}
  \label{fig:winner}
\end{figure}

\begin{table}[t]
  \centering
  \caption{Comparison between the reconstructed images and the initial
    images. The best scores are the lowest. }
  \label{tab:scores}
  \medskip
  \begin{tabular}{cccc|c|c}
    \hline                         
               &LR-H  &LR-K  &MR-H  &Combined  Gray   &$H$ Cube\\
    \hline
    \hline
    BSMEM      & 6.6    & 5.9    &10.4   & 7.6   & 13.9\\
    SQUEEZE    & 7.9    & 5.9    &22.7   &12.2   &[not submitted]\\
    WISARD     &11.5    &11.4    &37.9   &20.3   &47.0\\
    PPR        &40.0    &31.8    &42.1   &38.0   &45.6\\
    \hline
  \end{tabular}
\end{table}
To compare the reconstructed images which were submitted and the
initial data cube images, we used the following procedure:
\begin{enumerate}
\item Align all images using the centroid of a Gaussian fitted to the
  main star.  The image should not be Gaussian but this gives a good
  alignment.
\item Interpolate submitted images to the grid of the model images.
\item Model images were convolved by a 0.4\,mas circular Gaussian to
  get the resolution a little closer to that of the submissions.
\item Normalize all planes in all images such that the sum of the
  pixels in the box defined by corners [25,25] and [214,214] was 1.0.
  This includes the model images.
\item The comparison value for each image is $10^6$ times the RMS
  pixel-to-pixel difference between the submission and the model in
  the region defined by corners [25,25] and [214,214] plus the region
  defined by the corners [246,1013] and [266,1034] for each plane.
  These cover the regions of the primary star and the companion.  For
  the gray comparison, the final value is the average of those for the
  three images.  For the cube, the final value is the average over all
  planes.
\end{enumerate}
The best scores are the lowest. They are reported in Table
\ref{tab:scores}. The first two contests have been won by the BSMEM
entry and the last one by MIRA. While the top two contestants have
very similar results, the BSMEM entry represented by F.~Baron has
achieved the lowest score and is thereby declared the winner of this
year's contest (see Fig.~\ref{fig:winner}).

\section{Conclusion}

Some of the participating algorithms were able to perform relatively
faithful image reconstruction on the combination of a resolved stellar
surface and an unresolved faint companion. The new type of data set
consisting in spectral cubes was a challenge for all the contestants,
mostly because the computation time became an important issue. There
was no global image reconstruction, just one spectral channel by one
spectral channel reconstruction and no use of differential quantities. 
There were no attempts to get a spectrum of the companion relative to
the star so far. The organizer also hope that this type of exercize
will allow the different teams to test their the algorithms in order to
accomodate the full information contained in spectral data and improve
their faithfulness.

\acknowledgments     

Work by PRL was undertaken at the Jet Propulsion Laboratory, California Institute of Technology, under contract with the National Aeronautics and Space Administration
Work by WDC was supported by the National Radio Astronomy Observatory, a facility of the National Science Foundation operated under cooperative agreement by Associated Universities, Inc. 


\bibliography{2010-beauty-contest-spie}   

\begin{thebibliography}{10}

\bibitem{2004SPIE.5491..886L}
{Lawson}, P.~R., {Cotton}, W.~D., {Hummel}, C.~A., {Monnier}, J.~D., {Zhao},
  M., {Young}, J.~S., {Thorsteinsson}, H., {Meimon}, S.~C., {Mugnier}, L.~M.,
  {Le Besnerais}, G., {Thiebaut}, E.~M., and {Tuthill}, P.~G., ``{An
  interferometry imaging beauty contest},'' {\em Society of Photo-Optical
  Instrumentation Engineers (SPIE) Conference Series} {\bf 5491},  886 (2004).

\bibitem{2006SPIE.6268E..59L}
{Lawson}, P.~R., {Cotton}, W.~D., {Hummel}, C.~A., {Baron}, F., {Young}, J.~S.,
  {Kraus}, S., {Hofmann}, K., {Weigelt}, G.~P., {Ireland}, M., {Monnier},
  J.~D., {Thi{\'e}baut}, E., {Rengaswamy}, S., and {Chesneau}, O., ``{2006
  interferometry imaging beauty contest},'' {\em Society of Photo-Optical
  Instrumentation Engineers (SPIE) Conference Series} {\bf 6268} (2006).

\bibitem{2008SPIE.7013E..48C}
{Cotton}, W., {Monnier}, J., {Baron}, F., {Hofmann}, K., {Kraus}, S.,
  {Weigelt}, G., {Rengaswamy}, S., {Thi{\'e}baut}, E., {Lawson}, P., {Jaffe},
  W., {Hummel}, C., {Pauls}, T., {Schmitt}, H., {Tuthill}, P., and {Young}, J.,
  ``{2008 imaging beauty contest},'' {\em Society of Photo-Optical
  Instrumentation Engineers (SPIE) Conference Series} {\bf 7013} (2008).

\bibitem{2009A&A...506.1351C}
{Chiavassa}, A., {Plez}, B., {Josselin}, E., and {Freytag}, B., ``{Radiative
  hydrodynamics simulations of red supergiant stars. I. interpretation of
  interferometric observations},'' {\em \aap}~{\bf 506},  1351--1365 (2009).

\bibitem{2005ApJ...628..973L}
{Levesque}, E.~M., {Massey}, P., {Olsen}, K.~A.~G., {Plez}, B., {Josselin}, E.,
  {Maeder}, A., and {Meynet}, G., ``{The Effective Temperature Scale of
  Galactic Red Supergiants: Cool, but Not As Cool As We Thought},'' {\em
  \apj}~{\bf 628},  973--985 (2005).

\bibitem{2009A&A...498..321C}
{Chelli}, A., {Duvert}, G., {Malbet}, F., and {Kern}, P., ``{Phase closure
  nulling. Application to the spectroscopy of faint companions},'' {\em
  \aap}~{\bf 498},  321--327 (2009).

\bibitem{2002SPIE.4844..295D}
{Duvert}, G., {B{\'e}rio}, P., and {Malbet}, F., ``{ASPRO, a software to
  prepare observations with optical interferometers},'' {\em SPIE}~{\bf 4844},
  295 (2002).

\bibitem{2005PASP..117.1255P}
{Pauls}, T.~A., {Young}, J.~S., {Cotton}, W.~D., and {Monnier}, J.~D., ``{A
  Data Exchange Standard for Optical (Visible/IR) Interferometry},'' {\em
  \pasp}~{\bf 117},  1255--1262 (2005).

\bibitem{2004SPIE.5491..909M}
{Meimon}, S.~C., {Mugnier}, L.~M., and {Le Besnerais}, G., ``{A novel method of
  reconstruction for weak-phase optical interferometry},'' {\em Society of
  Photo-Optical Instrumentation Engineers (SPIE) Conference Series} {\bf 5491},
   909 (2004).

\bibitem{2008JOSAA..26..108M}
{Meimon}, S., {Mugnier}, L.~M., and {Le Besnerais}, G., ``{Self-calibration
  approach for optical long-baseline interferometry imaging},'' {\em Journal of
  the Optical Society of America A}~{\bf 26},  108 (2008).

\bibitem{1981MNRAS.196.1067C}
{Cornwell}, T.~J. and {Wilkinson}, P.~N., ``{A new method for making maps with
  unstable radio interferometers},'' {\em \mnras}~{\bf 196},  1067--1086
  (1981).

\bibitem{2005JOSAA..22.2348M}
{Meimon}, S., {Mugnier}, L.~M., and {Le Besnerais}, G., ``{Convex approximation
  to the likelihood criterion for aperture synthesis imaging},'' {\em Journal
  of the Optical Society of America A}~{\bf 22},  2348--2356 (2005).

\bibitem{1998JOSAA..15..811L}
{Lannes}, A., ``{Weak-phase imaging in optical interferometry},'' {\em Journal
  of the Optical Society of America A}~{\bf 15},  811--824 (1998).

\bibitem{1998ApOpt..37.4614C}
{Conan}, J., {Mugnier}, L.~M., {Fusco}, T., {Michau}, V., and {Rousset}, G.,
  ``{Myopic deconvolution of adaptive optics images by use of object and
  point-spread function power spectra},'' {\em \ao}~{\bf 37},  4614--4622
  (1998).

\bibitem{2008ISTSP...2..767L}
{Le Besnerais}, G., {Lacour}, S., {Mugnier}, L.~M., {Thiebaut}, E., {Perrin},
  G., and {Meimon}, S., ``{Advanced Imaging Methods for Long-Baseline Optical
  Interferometry},'' {\em IEEE Journal of Selected Topics in Signal Processing,
  Vol.~2, Issue 5, p.767-780}~{\bf 2},  767--780 (2008).

\bibitem{2008A&A...485..561L}
{Lacour}, S., {Meimon}, S., {Thi{\'e}baut}, E., {Perrin}, G., {Verhoelst}, T.,
  {Pedretti}, E., {Schuller}, P.~A., {Mugnier}, L., {Monnier}, J., {Berger},
  J.~P., {Haubois}, X., {Poncelet}, A., {Le Besnerais}, G., {Eriksson}, K.,
  {Millan-Gabet}, R., {Ragland}, S., {Lacasse}, M., and {Traub}, W., ``{The
  limb-darkened Arcturus: imaging with the IOTA/IONIC interferometer},'' {\em
  \aap}~{\bf 485},  561--570 (2008).

\bibitem{2009A&A...508..923H}
{Haubois}, X., {Perrin}, G., {Lacour}, S., {Verhoelst}, T., {Meimon}, S.,
  {Mugnier}, L., {Thi{\'e}baut}, E., {Berger}, J.~P., {Ridgway}, S.~T.,
  {Monnier}, J.~D., {Millan-Gabet}, R., and {Traub}, W., ``{Imaging the spotty
  surface of Betelgeuse in the H band},'' {\em \aap}~{\bf 508},  923--932
  (2009).

\bibitem{2009ApJ...707..632L}
{Lacour}, S., {Thi{\'e}baut}, E., {Perrin}, G., {Meimon}, S., {Haubois}, X.,
  {Pedretti}, E., {Ridgway}, S.~T., {Monnier}, J.~D., {Berger}, J.~P.,
  {Schuller}, P.~A., {Woodruff}, H., {Poncelet}, A., {Le Coroller}, H.,
  {Millan-Gabet}, R., {Lacasse}, M., and {Traub}, W., ``{The Pulsation of
  {$\chi$} Cygni Imaged by Optical Interferometry: A Novel Technique to Derive
  Distance and Mass of Mira Stars},'' {\em \apj}~{\bf 707},  632--643 (2009).

\end{thebibliography}
\bibliographystyle{spiebib}   

\end{document}